\documentclass[sigplan,nonacm]{acmart}

\usepackage{comment}
\newcommand{\sys}{\textit{Obsidian}}

\usepackage{tikz}
\usepackage{subcaption}
\usepackage{graphicx}
\usepackage{booktabs}
\usepackage{pgfplots}
\usepackage{pgfplotstable}
\usepackage{subcaption}
\usepackage{listings}
\usepackage{xcolor}
\usepackage{tcolorbox}
\tcbuselibrary{skins}
\usepackage{lipsum}
\usepackage{multirow}
\usepackage[linesnumbered,ruled]{algorithm2e}

\usepackage[]{hyperref}
\usepackage[capitalise,nameinlink]{cleveref}
\crefname{section}{\S}{\S}
\Crefname{section}{\S}{\S}

\newcommand{\etal}[1]{{#1}~{\it et~al.}}
\newcommand{\sarbartha}[1]{\textcolor{red}{[TODO: {#1}]}}

\newcommand*\circled[1]{\tikz[baseline=(char.base)]{
            \node[shape=circle,fill,inner sep=1pt] (char) {\textcolor{white}{#1}};}}
\def\BibTeX{{\rm B\kern-.05em{\sc i\kern-.025em b}\kern-.08em
    T\kern-.1667em\lower.7ex\hbox{E}\kern-.125emX}}   

\SetKwComment{Comment}{/* }{ */}
\RestyleAlgo{ruled}

\title{\sys{}: Cooperative State-Space Exploration for performant inference on secure ML accelerators}

\author{Sarbartha Banerjee}
\email{sarbartha@utexas.edu}
\affiliation{
  \institution{The University of Texas at Austin}
  \city{Austin}
  \state{Texas}
  \country{USA}
}

\author{Shijia Wei}
\email{shijiawei@utexas.edu}
\affiliation{
  \institution{The University of Texas at Austin}
    \city{Austin}
  \state{Texas}
  \country{USA}
}

\author{Prakash Ramrakhyani}
\email{prakash.ramrakhyani@arm.com}
\affiliation{
  \institution{ARM Inc.}
    \city{Austin}
  \state{Texas}
  \country{USA}
}

\author{Mohit Tiwari}
\email{tiwari@austin.utexas.edu}
\affiliation{
  \institution{The University of Texas at Austin}
    \city{Austin}
  \state{Texas}
  \country{USA}
}

\begin{document}
\begin{abstract}

Trusted execution environments (TEEs) for machine learning accelerators 
are indispensable in secure and efficient ML inference.
Optimizing workloads through state-space exploration for the accelerator architectures
improves performance and energy consumption.
However, such explorations are expensive and slow due to the large search space.
Current research has to use fast analytical models that forego critical hardware details and cross-layer opportunities
unique to the hardware security primitives.
While cycle-accurate models can theoretically reach better designs,
their high runtime cost restricts them to a smaller state space.

We present \sys{}, an optimization framework for finding the optimal mapping from ML kernels to a secure ML accelerator.
\sys{} addresses the above challenge by exploring the state space using analytical and cycle-accurate models cooperatively.
The two main exploration components include:
\textbf{(1)} A secure accelerator analytical model, that includes the effect of secure hardware while traversing the large mapping state space and produce the best $m$ model mappings;
\textbf{(2)} A compiler profiling step on a cycle-accurate model, that captures runtime bottlenecks to further improve execution runtime, energy and resource utilization and find the optimal model mapping. 

We compare our results to a baseline secure accelerator, comprising of the state-of-the-art security schemes obtained from guardnn~\cite{GuardNN2022DAC} and sesame~\cite{sesame}.
The analytical model reduces the inference latency by $20.5\%$ for a cloud and $8.4\%$ for an edge deployment with an energy improvement of $24\%$ and $19\%$ respectively.
The cycle-accurate model, further reduces the latency by $9.1\%$ for a cloud and $12.2\%$ for an edge with an energy improvement of $13.8\%$ and $13.1\%$.

\end{abstract}
\maketitle
\pagestyle{plain}
\section{Introduction}


Machine learning (ML) has become ubiquitous over the last decade 
with its ability to solve real-life problems like image classification~\cite{alexnet,resnetv2},
speech recognition~\cite{deepspeech}, language translation~\cite{translation}, playing games~\cite{AlphaGoZero}, recommendation systems~\cite{DLRM} and content generation~\cite{gpt3}.
Many of these applications use proprietary models, to infer sensitive data like 
genetic synthesis~\cite{Genetic_data_breach}, disease classification~\cite{medical_1,medical_2,medical_3},
face recognition~\cite{face_recog_1,face_recog_2}.
Prior secure accelerators~\cite{GuardNN2022DAC, mgx2022isca,keystone,securator} add an authenticated encryption (\textit{crypto}) block at the memory interface of the accelerator to protect data confidentiality and integrity in the main memory and while data transfer.

The memory protections, however, does not protect against model weight extraction attacks~\cite{model_weight_extraction_1,model_weight_extraction_2}.
These side-channel attacks, observe the memory access patterns~\cite{ReverseCNN}, or layer-wise memory bandwidth variations~\cite{DeepSniffer, band_util} to infer the layer hyperparameters.
Sesame~\cite{sesame} proposed to incorporate a constant traffic shaping logic (\textit{shaper}) to obfuscate the memory demand bandwidth, similar to CPU traffic shaping mitigation~\cite{dagguise,camouflage}.
It also proposes a hardware zeroization block (\textit{zeroizer}) to clean secret data from scratchpads before context switch to protect memory safety.

While designs like Sesame incorporate side-channel defences in ML accelerators,
they introduce a significant latency and power overhead beyond secure accelerators with 
only memory protections.
Such overhead prevents side- channel defense techniques
from getting adopted in real-time edge ML inferences~\cite{alexa, edgeTPU, eyeriss, jetson} and cloud machine-learning-as-a-service (MLaaS) deployments~\cite{aws_ml_marketplace,tpuv4,vaswani2023confidential}.
Data-centric approaches like tensor tiling,
loop ordering and dataflow scheduling~\cite{datacentric,timeloop,maestro}
improve the inference latency by restructuring tensor computations
for non-secure ML accelerators, but present a large state space.
Several optimization techniques like genetic algorithm~\cite{gamma,digamma}, search-space pruning~\cite{timeloop} and constrained optimization~\cite{cosa} is used to traverse the search space. 
However, incorporating the security primitives (i.e. \textit{crypto, shaper and zeroizer})
into the analytical cost model causes search space explosion because these
features entitle a spectrum of configurations.
A second technique to improve the inference latency in secure ML accelerators,
is to perform profile-guided optimizations~\cite{AutoTVM} via compilers. 
While, compiler profiling yields better mapping results, as it is run either in real hardware~\cite{VTA} or cycle-accurate simulator~\cite{scalesim}, 
a higher runtime cost restricts the state exploration.

Beyond state-space exploration, the inference latency can be further improved with faster secure block implementations and using domain knowledge. 
For instance, secureloop~\cite{secureloop} used pipelined~\cite{aespipe} or parallel~\cite{aespll} \textit{crypto} block implementations to increase throughput.
MGX~\cite{mgx2022isca} used domain knowledge like the read-only nature of model weights eliminates the integrity counters.
Moreover, the integrity counters for the input and output feature maps is generated inside the accelerator to reduce metadata memory traffic. 
Finally, sesame~\cite{sesame} used explicit instructions to perform zeroization of only secret scratchpad data.

In this work, we introduce \sys{}, a design-space exploration tool to improve inference latency for secure ML accelerators. 
\sys{} build an analytical model for the secure blocks (\textit{crypto, shaper and zeroizer}) to find a group of best candidate mappings;
The analytical model first uses gamma~\cite{gamma} mapping tool to find the \textit{top-k} mapping of each layer.
The tool then finds 
\textbf{(1)} The optimal integrity hash granularity for each layer;
\textbf{(2)} Estimates the \textit{shaper} bandwidth to maximize the energy-delay product;
\textbf{(3)} Evaluates the context-switch latency based on scratchpad utilization.
With all these inputs, a simulated annealing algorithm is run to reduce the search space to \textit{top-m} model mappings.

Next, it perform compiler profiling with a cycle-accurate simulator to  to efficiently utilize the \textit{shaper} and the \textit{zeroizer}.
The \textit{shaper} introduces pipeline stalls during high bandwidth data transfers, while adding fake transactions during idle times to obfuscate the memory demand traffic.
The profiler efficiently schedule data transfers, even across layer boundaries to reduce system stalls, while reciprocating fake transactions by real ones. 
The \textit{zeroizer} introduces additional cycles during context switch in time-shared multi-tenant accelerators.
The profiler introduces a data reuse analysis to proactively clear scratchpad data, which are no longer needed for future computation, amortizing the context switch latency.
The analytical stage efficiently searches a large state-space, while the profiling phase addresses runtime bottlenecks.




We evaluate \sys{} secure state-space exploration framework in two scenarios: \textbf{(1)} A cloud ML-as-a-Service (\textit{cloud})
and \textbf{(2)} An edge ML inference device (\textit{edge}).
The analytical phase reduces the inference latency by $20.5\%$ in \textit{cloud} and $8.4\%$ in \textit{edge}, 
with $24\%$ and $19\%$ less energy consumption, compared to a baseline secure accelerator with \textit{crypto, shaper} and \textit{zeroizer} blocks.
Next, the compiler profiling, further reduces the latency by $9.1\%$ for \textit{cloud} and $12.2\%$ for \textit{edge}, with 
$13.8\%$ and $13.1\%$ energy reduction. 
Finally, the proactive \textit{zeroizer} reduces the data volume by $22.8\%$ in \textit{cloud} and $11.2\%$ in \textit{edge} configurations, if the model is context-switched after each layer 
in a time-shared multi-tenant deployment.

\vspace{-1em}

\section{Background}

\subsection{State space exploration of ML workloads}
\label{subsec:bg_dse}
\begin{figure*}[t]
    \centering
    \hspace{-1em}
    \begin{subfigure}{0.5\linewidth}
        \includegraphics[width=0.9\linewidth]{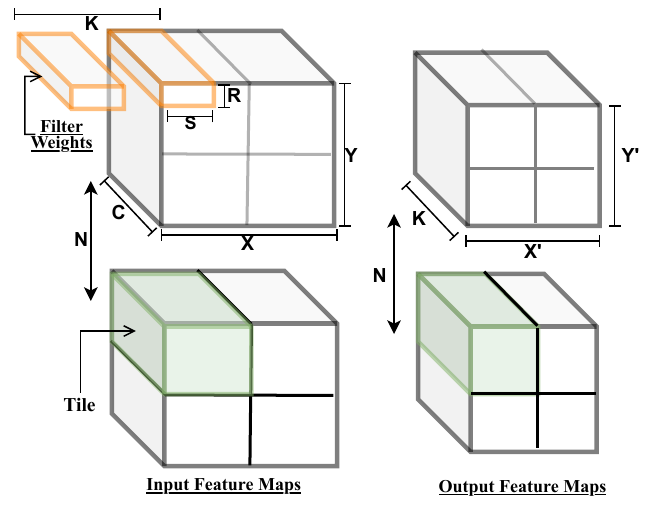}
        \caption{\textbf{The components of a convolution operation.}}
        \label{fig:dnn_comp}
    \end{subfigure}
    \begin{subfigure}{0.5\linewidth}
        \includegraphics[width=\linewidth]{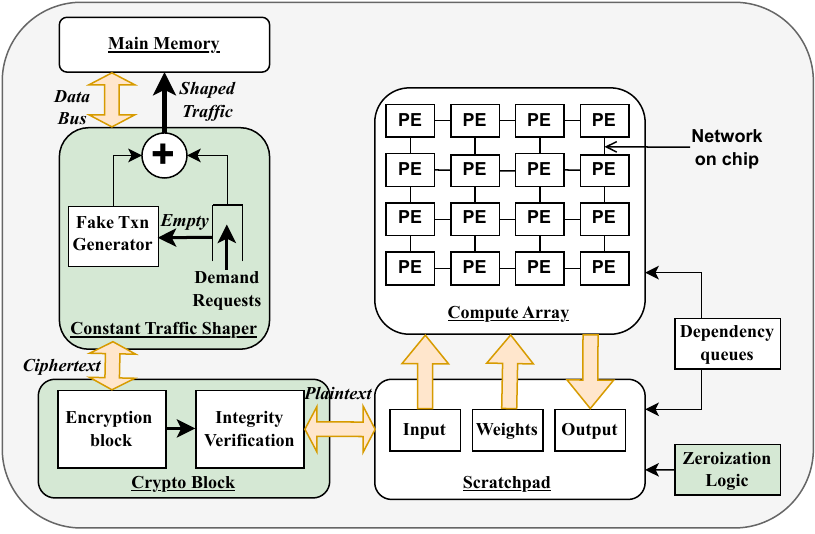}
        \caption{\textbf{Secure ML inference accelerator architecture}}
        \label{fig:sec_arch}
    \end{subfigure}
    \vspace{-1em}
    \caption{\textbf{Design space exploration is performed by ordering and tiling tensor dimensions. The optimal mapping is executed in a secure accelerator. 
    The green blocks show the secure hardware, while datapath is shown in yellow. }}
    \vspace{-1em}
    \label{fig:figure1}
\end{figure*}

The state space of each layer of an ML workload spans across different tile size, dataflow and loop ordering mappings~\cite{datacentric}. 
~\cref{fig:dnn_comp} shows the convolution layer, a widely used operation in deep neural networks.
$K$ model filter weights, each of dimension $R\times S$ is multiplied with a set of $N$ input feature maps (\textit{ifmap}) with dimension $X\times Y$
to create $N$ output feature maps (\textit{ofmap}), each of size $K\times X' \times Y'$. The $X'$ and $Y'$ depend on $X,Y,R,S$ and the filter stride.
This operation creates a nested loop of seven independent dimensions ($K,R,S,N,X,Y,C$), referred to as loop-nest in literature~\cite{timeloop}.
These seven loops can be re-ordered to find the best data reuse, creating a search space of $7!$.
Each of the dimension can be tiled to extract locality during computation.
While tiling provides more reuse opportunities, it increases the state space to several hundreds of thousands of mappings.
Prior work~\cite{gamma,mindmappings,digamma} used random search, genetic algorithms, machine learning etc. to search such a large space to find an optimal mapping with the best latency or energy consumption.
These mappings are evaluated with a hardware cost model~\cite{maestro,timeloop}.
The cost model contains the power, area and latency parameters for each accelerator component like the processing unit (PE), 
scratchpad sram buffers, network and off-chip bandwidth. 
It additionally reports energy with activity counts for each component to compute energy consumption.

AutoTVM~\cite{AutoTVM} takes a different approach of performing the tile search through compiler profiling.
The different mappings are run in a real hardware~\cite{VTA} or a cycle-accurate simulator~\cite{tvm}, instead of an analytical cost model. 
It searches a much smaller design space (ideal for small accelerators), but is able to factor in system bottlenecks in the mapping phase.


\vspace{-1em}
\subsection{ML inference accelerator architecture}
\label{subsec:bg_ml_design}
ML accelerators are specialized dataflow architectures performing low latency inference. 
The popular accelerators~\cite{TPU,VTA,maestro,eie} have a decoupled-access execute (DAE) architecture~\cite{DAE} with their own instruction set using systolic arrays~\cite{systolic} for efficient data movement.
It comprises a compute array of processing elements (PE), to perform matrix-matrix or matrix-vector computations as shown in ~\cref{fig:sec_arch}. 
The output of a compute operation is forwarded to the next CU through a network-on-chip (NoC).
There are individual \textit{scratchpads} for data inputs, model weights, and inference outputs. 
These \textit{scratchpads} are double-buffered to perform data movement (loads and stores) and computation in parallel. 
\textit{Dependency queues} between the scratchpads and the compute blocks handle data dependencies.
The co-processor (\textit{CPU}) interacts with the model vendor and the data owner to load data into the respective memory regions. 
The model algorithm is loaded into an instruction queue.
Certain ML accelerators have a dedicated ISA with binary generated in a specialized compiler~\cite{tvm,mlir}.
Explicit \texttt{load} and \texttt{store} instructions perform data transfer, while \texttt{gemm} and \texttt{alu} instructions perform tensor computation.
The compute blocks either accept data from the scratchpad buffer or from the adjacent compute blocks in a systolic pattern~\cite{maestro, tpuv4}. 
\vspace{-1em}
\subsection{Secure hardware blocks in an ML accelerator}
\label{subsec:bg_sec_hw}
The green boxes in ~\cref{fig:sec_arch} show the security blocks.
A constant memory traffic shaper (\textit{shaper}) includes a demand queue to hold pending data transfer requests, 
and dispatches them to the main memory at fixed intervals.
The instruction queue is stalled if there the demand queue is full, while a fake transaction is generated into the same bank if the demand queue is empty.
The \textit{shaper} obfuscates the memory traffic observation by maintaining a constant throughput.

Secret data encryption and integrity verification is done in the crypto block.
Secret data is stored as ciphertext in main memory along with a data hash.
The crypto block converts it to plaintext, and verifies the data hash, before loading data in respective scratchpads.
MGX~\cite{mgx2022isca} proposed to generate integrity counters inside the accelerator, eliminating the need for merkle tree 
traversal and additional memory transactions.


To protect memory safety against use-after-free attacks in a multi-tenant environment, 
sesame further proposes a zeroization logic to clear secret data from the scratchpad during context switch.

The goal of this paper is to model these security hardware blocks to find the optimal hash granularity for the \textit{crypto} block, 
the optimal memory bandwidth for the \textit{shaper} to reduce the overall energy consumption, 
and to reduce the zeroization overhead during context switch.
and to increase the overall resource utilization by reducing the number of fake transactions in the traffic shaper.
\vspace{-1em}
\subsection{Context switch in multi-tenant accelerators}
\label{subsec:bg_cs}
Time-sharing multiple ML model execution is critical in providing an overall quality-of-service (QoS) for ML accelerators. 
Prior research~\cite{prema, planaria} schedule multiple DNN workloads temporally or spatially in a single accelerator. 
The context switch (CS) is typically done at a layer boundary, and depends on the deadline of the pending workloads. 
The entire scratchpad is zeroized to protect memory safety violations.
Sesame clears only a subset of scratchpad locations that hold secret data with a hardware \textit{zeroizer}.
This not only improves CS latency, but also reduces the system energy consumption.
The goal of \sys{} is to interleave the zeroize operation with computation, to further minimize the CS latency.

\section{Motivation and Goals}

\subsection{Execution bottlenecks of a secure ML accelerator}
\label{subsec:moti_ineff}
The constant memory traffic shaper proposed by sesame provides a fixed memory bandwidth for the entire model. 
While it obfuscates the memory utilization side-channel, a lot of fake transactions are generated, especially for the compute-bound layers.
Reducing the traffic shaper bandwidth reduces the number of fake transactions, but also throttles larger layer tiles, delaying the execution. 
Mapping space exploration (MSE) can find optimal layers, with a greater synergy between the data transfer and the compute units.

The data throughput to the accelerator depends on the \textit{crypto} block as each data needs decryption and integrity verification. 
A good MSE should incorporate the \textit{crypto} throughput, while finding the optimal mapping. 
The \textit{crypto} block performs additional memory transactions to load the hash metadata from memory. 
This can be done by increasing the data granularity for hash calculation. 
A large granularity, however, can lead to additional redundant data loads, just for integrity verification, giving diminishing returns.
The choice of hash granularity is workload dependent.
\sys{} explores across multiple data sizes for each layer to find an optimal integrity granularity.

The context switch overhead in a multi-tenant setting is the next bottleneck. 
Context switching additionally involves removing secret data and states from on-chip scratchpad, buffers and queues.

\subsection{Challenges in design space exploration of secure ML accelerators}
Finding the optimal mapping for each layer is critical to reduce the inference latency, and increase the effective accelerator resource utilization. 
Analytical models~\cite{maestro,timeloop} are widely used for design space exploration of non-secure ML accelerators as discussed in ~\cref{subsec:bg_dse}.
Since the \textit{crypto, shaper} and \textit{zeroizer} directly impacts the execution critical path, the analytical model should include these components.
This introduces additional tunable parameters, like data hash granularity in \textit{crypto} block, optimal bandwidth for the \textit{shaper} etc.,
which further increases the search space.

Another challenge is that, many of these secure hardware units have low level optimizations, that are not captured by the analytical model. 
For instance, the vanilla constant traffic shaping logic used in sesame, generates a lot of fake transactions as discussed in ~\cref{subsec:moti_ineff}.
Data from high throughput layers, later in the model can benefit from idle memory cycles of previous smaller layers.
Analyzing such cross-layer optimizations is infeasible in an analytical model for large models like the transformers.
The analytical model also lacks cycle-accurate information of memory idle cycles and scratchpad utilization, which is required for exploring these load-sharing optimizations.

While compiler profiling has the runtime resource utilization, it cannot search such a large state space due to its higher runtime.
The goal of \sys{} is to harness the benefits of both the analytical model and the compiler profiling. 
Hence, we take a cooperative approach in finding the optimal mapping. 
The analytical model reduces the search space to find near-optimal mappings, 
which are profiled in a compiler to extract cross-layer load-sharing opportunities to find the optimal mapping.

\subsection{Effect of context switch overhead in multi-tenant MLaaS accelerators}
As mentioned in ~\cref{subsec:bg_cs}, sesame zeroizes each secret scratchpad location on a context switch. 
Given the large scratchpad size in cloud MLaaS deployments, zeroization increases the overall context switch time significantly, 
as the \texttt{zeroize} instruction is invoked after the completion of a layer computation.
This costs non-billing machine cycles to the cloud service provider (CSP).
\sys{} performs compiler analysis to identify scratchpad data that can be removed immediately after compute consumption, against flushing it during context switch.


\subsection{Threat model and assumptions}
An attacker can observe and tamper private data in the off-chip memory and during data transfer over the memory bus.
The attacker has further access to the utilizaton information of the read and write memory bus, and can use it to find private input size or infer confidential model structure. 
An attacker has the capability to read scratchpad locations, before and after a victim tenant execution in a multi-tenant ML accelerator.
The security guarantees are the same as sesame~\cite{sesame}, with the addition of the data integrity protection, as in guardnn~\cite{GuardNN2022DAC}.

Our focus is on the accelerator runtime, and we assume that the accelerator is bootstrapped with secure boot and do not contain any hardware trojans.
We assume that a tenant has verified the accelerator through remote attestation before transferring private model and input into the accelerator as demonstrated in prior works~\cite{sgx,shef}. 
The accelerator system should also have a protected storage to store secret keys for each tenant.

\section{\sys{} state space exploration}
\label{sec:dse}
\begin{figure*}[htbp]
    \centering
    \includegraphics[width=\linewidth]{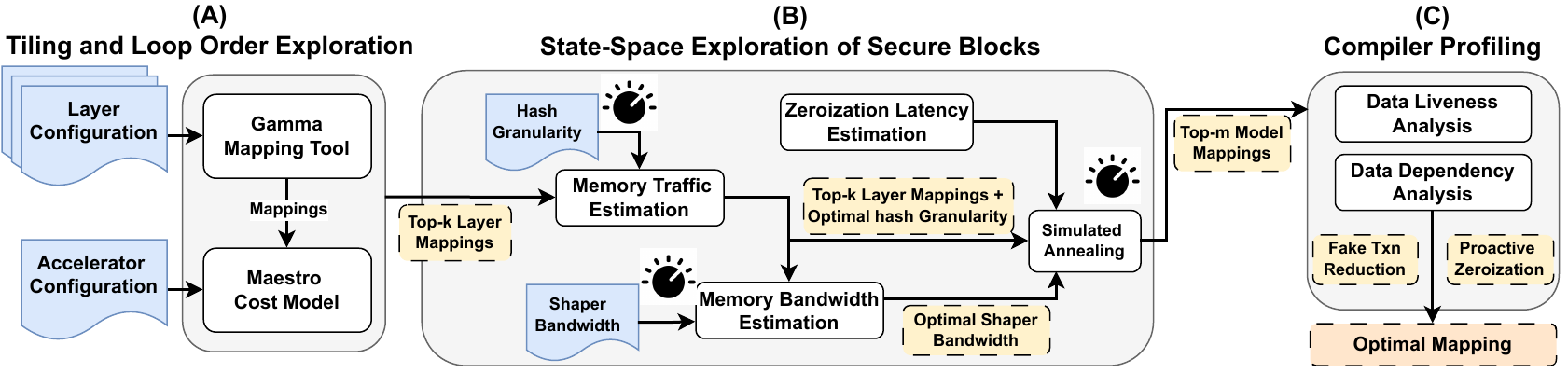}
    \caption{\textbf{The different optimization phases of \sys{}}}
    \vspace{-1em}
    \label{fig:analytical_model}
    \end{figure*}
    
\sys{} performs a cooperative mapping state space exploration for secure ML accelerators.
The first stage is an analytical phase, that finds the \textit{top-m} mappings for the workload, 
followed by a compiler profiling phase to perform cross layer data transfer to reduce execution stalls and increase resource utilization. 

We add the impact of \textit{crypto, shaper} and \textit{zeroizer} to maestro cost model, a widely used analytical model for mapping space exploration for non-secure ML accelerators.
In addition to finding the optimal tile size and loop order for each layer, the \sys{} model  
\textbf{(1)} Finds the optimal data granularity for hash calculation;
\textbf{(2)} Explores the optimal memory bandwidth for the \textit{shaper}; and
\textbf{(3)} The zeroization cost associated with each mapping for every layer.
Once, we have the \textit{top-k} layer mappings for each layer, 
\sys{} uses simulated annealing to find the \textit{top-m} mappings for the workload, reducing the state space further from $k^l \rightarrow m$ mappings.

These $m$ mappings are profiled in a compiler, which comprises of two dataflow passes:
\textbf{(1)} A data liveness analysis, performing data reuse analysis on consumed data to decide if it can be zeroized from the scratchpad.
This analysis also invalidates stale data in each scratchpad, making it available for new subsequent data, that can reciprocate fake transactions in the \textit{shaper}.
\textbf{(2)} A data dependency analysis finds the earliest availability of each \textit{ifmap}, that can be prefetched into the scratchpad. 

\subsection{\sys{} analytical model}
\label{subsec:dse_anal}

\subsubsection{ML layer computation mapping}
\label{subsec:dnn_anal}
~\cref{fig:analytical_model}(A) illustrates the steps of finding the \textit{top-k} optimal mapping for each DNN layer. 
We leverage the maestro cost model to estimate the runtime for each mappings, and use the gamma mapping tool to traverse the search space.
The maestro model takes hardware parameters as input, which includes the number of PEs, the design dataflow, the size of each scratchpad,
the NoC bandwidth, and the offchip memory bandwidth.
All these parameters are the same as a non-secure ML accelerator, except for the offchip memory bandwidth.
Each data traverses through a constant \textit{shaper} with a fixed memory bandwidth, followed by the decryption and integrity verification in the \textit{crypto} block. 
The effective memory bandwidth to the scratchpad effectively becomes the \textit{crypto} block throughput. 
The state space is traversed to find the \textit{top-k} mappings ($map_k$) for each layer.

The choice of $k$ defines the state space available to subsequent optimization opportunities. 
Choosing a high value of parameter k explodes the search space, while a low value minimizes further optimization opportunities. 
A good set of chosen mappings should have a small variance in inference latency and energy consumption, compared to the top mapping.
For a workload with $l$ layers, this step reduces the state space to $k^l$.
We reason the choice of $k$ in finding the optimal mapping in ~\cref{subsec:eval_anal_param}.

\subsubsection{Exploring the optimal integrity granularity}
\label{subsec:opt_hash_design}
Data is streamed from the main memory at tile granularity, which normally spans across several hundreds of bytes. 
Having an integrity verification at 64 byte granularity increases the number of hash metadata memory traffic. 
A larger data granularity of hash verification (\textit{AuthBlock}) reduces the number of metadata loads for certain layers.
However, there are inefficiencies associated with a larger \textit{AuthBlock}. 
Larger \textit{AuthBlock} might require redundant data loads, just to perform integrity verification, if only a subset of the \textit{AuthBlock} is required for computation. 
With a limited memory bandwidth available in the \textit{shaper}, making an optimal \textit{AuthBlock} choice is critical to DSE.
We make a key observation, that even when we load additional redundant data for a larger authentication block, 
many of them are reused later in the computation. 
We take this reuse opportunity into consideration while searching for the optimal \textit{AuthBlock} for each \textit{top-k} layer mapping.

The memory traffic estimation block in ~\cref{fig:analytical_model}(B) estimates the total memory traffic generated for each \textit{AuthBlock}.
For a workload with $l$ layers, $k$ mappings for each layer and $h$ possible \textit{AuthBlock} values,
the memory traffic estimation step is performed for $k \times l \times h$ times.

\SetKwComment{Comment}{/* }{ */}
\begin{algorithm}
\SetAlgoLined
\SetAlTitleFnt{Ariel}
\caption{Finding the optimal AuthBlock}
\label{alg:opt_auth}
k $\gets$ NumMapping \\
h $\gets$ AuthBlockChoices \\
traceSeq $\gets$ [] \\
memTraffic[k] $\gets$ 0 \\
$h_{opt} \gets 0$
\For{i = 1 to k} {
\For{j = 1 to h} { 
traceSeq = GenerateMemTrace ($map_i$, h) \\
$N_D$, $N_R$,$N_I$ = DataReuse (traceSeq) \\
traffic = $(N_D + N_R + N_I)$ \\
\If{memTraffic[i] > traffic} {
memTraffic[i] = traffic\\
$h_{opt}$ = j
}}}

\SetKwProg{Fn}{Function}{:}{}
\Fn{DataReuse (traceSeq)}{
\For{addr in traceSeq}
{
\If{addr.label == 'redundant'}
{
spad = addr.type \\
dist $\gets$ SearchDownstream(addr, spad)}
\If{dist < spadSize[spad]}
{addr.label = 'demand' \\
RemoveNextAccess(addr)
}
}
}
\end{algorithm}

~\cref{alg:opt_auth} details the steps of \textit{AuthBlock} exploration for each layer.
First, a memory trace is constructed for each mapping $i$ and granularity $j$ by the \texttt{GenerateMemTrace} function in line 7.
This function first creates a demand transaction address sequence. 
Next, it generates the hash integrity and redundant transactions and insert it adjacent to corresponding demand transaction to create a \texttt{traceSeq}.
Each element in \texttt{traceSeq} is a tuple of \textit{\{addr, type, label\}} fields.
The \textit{addr} field stores the memory address, the \textit{type} field denotes if the data is destined for \textit{ifmap, weight or ofmap} scratchpad, 
and the \textit{label} field indicates if the data is a demand, a redundant or a hash request. 
The \texttt{traceSeq} list is fed in the \texttt{DataReuse} function as shown in line 8.
The \texttt{DataReuse} function finds if any of the \textit{redundant} request is reused as a \textit{demand} request.
The \textit{label} of those redundant transaction is promoted to \textit{demand} and the corresponding demand request is removed.
A valid data reuse happens if the reuse distance is less than the scratchpad size as shown in lines 19-24.
Finally, memory traffic is calculated by adding the number of demand requests ($N_D$), redundant requests ($N_R$) and integrity requests ($N_I$) as shown in line 10. 
Lines 11-12 finds the hash granularity with the minimum number of memory transactions for the $i^{th}$ layer.



\subsubsection{Exploring the optimal \textit{shaper} bandwidth}
In this phase, we compute the \textit{shaper} bandwidth and the corresponding energy consumption for each $map_k$ with optimal granularity.
The scratchpad is double-buffered and data load of $tile_i$ occurs in parallel to the $tile_{i-1}$ computation. 
The mapping runtime ($R_i$) for a layer $i$ is calculated in ~\cref{subsec:dnn_anal} with the offchip memory bandwidth set as the \textit{crypto} throughput ($bw_{crypto}$). 
For a memory bound (MB) layer mapping, the optimal \textit{shaper} bandwidth ($bw_{shaper}$) is calculated as the minimum bandwidth required for the \textit{crypto} block to maintain the throughput of real data as shown in ~\cref{eq1}. 
The \textit{shaper} will additionally have to load $N_R$ and $N_I$ transactions for integrity verification. 
For a compute bound (CB) layer mapping, $bw_{crypto}$ can be relaxed to load $N_D$ demand transactions by the tile compute time, in addition to integrity traffic as shown in ~\cref{eq2}
\begin{equation}
\label{eq1}
     bw_{shaper}^{MB}= bw_{crypto} + (N_R +N_I)/R_i
\end{equation}
\vspace{-1em}
\begin{equation}
\label{eq2}
     bw_{shaper}^{CB}=  (N_D+N_R+N_I)/R_i
\end{equation}

\subsubsection{Finding the top-m model mappings}
So far, we generated the \textit{top-k} layer mappings with optimal hash granularity and \textit{shaper} bandwidth requirement. 
There is still a state space of $k^l$ for a model with $l$ layers which is quite large for performing compiler profiling. 
In this phase, we use simulated annealing to reduce the state space to \textit{top-m} mappings. 
Simulated annealing is a meta-heuristic algorithm and is designed to find optimal solutions for a large state space.

\textbf{Simulated annealing algorithm}:
The algorithm is detailed in ~\cref{alg:two}. 
The first step is to choose an initial model mappings ($map_i$).
A mapping from each layer is randomly chosen to serve as the initial mapping as shown in lines 4-6.
The energy-delay product (EDP) is calculated for this mapping by the \texttt{EnergyDelayCost} function in line 8.
Next, we run the simulated annealing algorithm for $N$ iterations. 
For each iteration, a new map is generated by the \texttt{GetNeighbor} function, and the EDP of the new mapping is calculated in line 13.
If the new map has a better EDP, it replaces the old mapping in the next iteration. 
However, if the old mapping was better, it can still be replaced by the new mapping given the temperature as shown in line 17.
The reason is to avoid local minimas while traversing the state space. 
Line 20 updates the temperature across iterations, which decides the probability of selecting a new mapping with a larger EDP than the previous mapping. 
The initial temperature is set high to enable larger exploration and is reduced to $T_{final}$ in the later iterations to converge on the optimal result. 
This algorithm is spawned $m$ times in parallel, each with a different initial mapping to find the \textit{top-m} layer mappings. 
Multiple runs with a different initial mapping provide higher probability of finding optimal solutions in different regions of the mapping space.

\SetKwComment{Comment}{//}{}
\begin{algorithm}
\SetAlgoLined
\SetAlTitleFnt{Ariel}
\caption{Simulated annealing for generating top-m from $k^l$ model mappings}
\label{alg:two}
L $\gets$ numLayers \\
N $\gets$ numIterations\\
$map_i \gets []$ \\
\Comment{Choose an initial model mappings}
\For{l $\gets$ 1 to L}
{
 $r \gets random(1,k)$ \\
 $map_i.append(map^l_k[r])$
}
\Comment{Optimal mapping with simulated annealing}
Cost $\gets$ EnergyDelayCost($map_i$) \\
t $\gets T_{init}$ \\
\For{n $\gets$ 1 to N}
{
 $i \gets random(1,n)$\\
 $NewMap_i \gets$ GetNeighbor($map_i$)\\
 NewCost  $\gets$ EnergyDelayCost($NewMap_i$)\\
 \If{NewCost < Cost}
 {
 $map_i = NewMap_i$
 }
 \ElseIf{$e^{(Cost - NewCost)/t}$ > random(0,1)}
 {
   $map_i = NewMap_i$
 }
 t $\gets$ UpdateTemperature($T_{init}, T_{final}$,n)
 }

\end{algorithm}

\textbf{Computing the objective function}:
The objective function used is energy-delay product (EDP) of each mapping as shown in ~\cref{fig:analytical_model}(B).
At each iteration $n$, the algorithm computes the \texttt{EnergyDelayCost} function to compare between the model mappings. 
A new model mapping $NewMap_i$ comprises of a set of chosen layer mappings. 
The model EDP is simply the sum of the EDP the individual layer mappings.
Each layer mapping is either categorized as a compute-bound (CB) or a memory-bound (MB) layer. 
The latency of a CB mapping is directly taken from the output of the tiling phase (~\cref{subsec:dse_anal}). 
The \textit{shaper} and the \textit{crypto} units do not introduce any additional delay, as the PE throughput is the bottleneck in this case. 
In a MB mapping, the system is bottlenecked by the \textit{crypto} unit. 
Hence, the runtime is computed by the multiplying the number of memory transactions and the shaper bandwidth. 
The energy is computed by the activity count of all the accelerator components.
The CB mapping also includes the energy spent on fake transactions sent to memory. 

\textbf{Equalizing the shaper bandwidth}
The different layers of the optimal mapping can have different bandwidth. 
But to ensure security, the \textit{shaper} should have a constant bandwidth across the entire model execution. 
So, while computing the \texttt{EnergyDelayCost} function, the memory bandwidth tuned to multiple values.
First, the model is run with the highest layer bandwidth to compute the energy. 
Then the bandwidth is lowered in fixed steps to iteratively find the optimal \textit{shaper} bandwidth for the entire model. 
\subsubsection{Secret zeroization}
\label{subsec:zero_dse}
If the state-space exploration is performed for a multi-tenant deployment, the zeroization latency should also be considered for each layer mapping. 
The zeroization latency for each mapping is calculated from the peak scratchpad utilization ($util^{s}_{peak}$ bits) reported by the tiling phase, 
where $s$ denotes the scratchpad type (\textit{ifmap, weight, ofmap}).
if the scratchpad srams are addressed by $a$ bits, and it takes $C$ cycle to write an address, the latency is given by ~\cref{eq4}. 
\begin{equation}
\label{eq4}
     t_{zero} = \Sigma_{s = i,w,o}\lceil util^{s}_{p}/a \rceil * C
\end{equation}
This latency is added to the mapping latency of each layer.
The energy consumption is computed by the same equation, by multiplying the unit sram write energy multiplied by the peak scratchpad utilization.
\subsubsection{Impact of exploration parameters}
\label{subsec:eval_anal_param}
Our state space exploration consist of two parameters: 
\textbf{(1)} The number of layer mappings ($k$) generated by the tiling exploration phase,
and \textbf{(2)} The number of model mappings ($m$) generated by the simulated annealing algorithm.

A large $k$ increases the state space of the simulated annealing, and might result in poor solution. 
It also increases the time to compute optimal hash and shaper bandwidth. 
On the other hand, reducing the value of $k$ yield a shallow search space.
For each layer, we discard mappings that have >$10\%$ overhead from the best mapping. 
We observe that these mappings are rarely chosen in the simulated annealing step. 
Moreover, we retain more mappings for layers having a higher runtime, like the embedding layer in transformer.  
Since, these layers contribute significantly to the overall inference latency, enabling the model mapper to 
explore more options often lead to better solutions. 
We chose a variable $k$ ranging from 2 for small layers to 16 for the transformer embedding layer 
using a simple heuristic, that is a function of layer runtime and the number of available mappings.

The model mapping parameter is critical to determine the time taken for the compiler profiling. 
The value of m is varied for the most complex model in our evaluation (transformer) 
and we see that the performance improvement from the profiling stage flattens out at m=40. 
So, we use this value for all our benchmarks.

\subsection{\sys{} compiler profiling}
The analytical state space exploration deduces the \textit{top-m} model mappings.
This provides the compiler profiling step with a feasible search-space to find the optimal mapping.
In this phase, \sys{} opportunistically re-orders the memory transactions to utilize the fake transaction intervals.
This stage is particularly useful when consecutive layers are CB and MB.
The memory idle cycles of the CB layer can be used by the MB layer to prefetch data, reducing the memory stalls. 
The profiles performs a \textbf{(1)} Data liveness analysis to minimize scratchpad occupancy at each cycle, by only retaining active data; 
and a \textbf{(2)} Data dependency analysis to find demand addresses that are ready to be promoted in the instruction stream.
The output of this compiler stage is the optimal mapping of the entire model, that is used during runtime execution.

\subsubsection{Data liveness analysis}
Prefetching new data from subsequent layers depends on the scratchpad availability. 
This compiler pass performs reuse analysis on the data consumed by the compute unit at each cycle.
If the data has a large reuse distance, greater than the scratchpad size, the corresponding scratchpad location is added to a free list. 
Future demand requests are promoted up the instruction stream to fill the scratchpad. 
Once, they are added to the demand queue, the \textit{shaper} loads these data instead of sending fake transactions to the main memory. 
If there are multiple candidates can be loaded at any instant, they are sorted based on the computation sequence.  
This ensures better utilization by only maintaining active variables in the scratchpad for the minimal time. 
This analysis is applied to all the top-m layer mappings, which are then ranked based on runtime.

This pass is also used to perform proactive zeroization.
If there is a context switch scheduled at the end of a layer, instead of loading new demand requests from later layers, 
the scratchpad location is zeroized immediately after compute consumption.
This process is interleaved with the layer computation, and reduces zeroization data volume during context switch.

Note that the data prefetching approach can also be applied in loading integrity metadata. 
If the design has a hash storage buffer, memory idle cycles can be used to load future data hash values.
This increases the integrity throughput for layers, which has a low integrity granularity (e.g., depthwise convolution) and a high metadata traffic.
We leave this optimization to future work.

\subsubsection{Data dependency analysis}
The prefetched data need to be ready, before it is being brought into the scratchpad.
A data dependency analysis is performed by the compiler to ensure correctness of prefetched loads, as \sys{} changes the instruction order.
This is not required for loading model weights, as they are read-only and does not depend on past layers.
Input feature map loads, on the other hand, can only be prefetched if the output feature maps are generated from the last layer. 
We perform a dataflow analysis to find the feature map dependency between layers and create a sequence of available data. 
The output correctness is ensured even after the change in the instruction order. 

The data dependency analysis also open the opportunity for store to load forwarding. 
If a freshly generated \textit{ofmap} will be immediately used as \textit{ifmap} for a subsequent layer, 
the data is directly forwarded from output to input scratchpad. 
Store to load forwarding not only reduces the memory traffic congestion, but also eliminates the \textit{crypto} delay.
This is particularly useful for small layers used in resnet and transformer models. 
The smaller layers have less scratchpad occupancy, leaving ample opportunity of store to load forwarding.
Overall, we perform this analysis for all the top-m configurations to generate the instruction stream for the optimal mapping sequence. 

\vspace{-1em}
\section{Implementation}
\label{sec:Implementation}
First, we enlist the secure accelerator configurations used in our evaluation in ~\cref{subsec:acc_config}.
Next, we describe the different components of the analytical model in ~\cref{subsec:impl_anal}, 
and the cycle-accurate simulator in ~\cref{subsec:impl_sim}.
Finally, we list the benchmarks used in our evaluation in ~\cref{subsec:impl_bench}.
\vspace{-0.5em}
\subsection{Accelerator configuration}
\label{subsec:acc_config}
 \begin{table}[b]
    \centering
    \resizebox{\linewidth}{!}{%
    \begin{tabular}{|l|cc|}
        \hline
        \textbf{Parameter} &
        \textbf{Cloud MLaaS} & 
        \textbf{Edge Device} \\
        \hline
        Dataflow & \multicolumn{2}{c|}{ output-stationary (os)} \\
        Memory type & DDR4 @ 2400MHz  & DDR3 @ 1066MHz \\
        Crypto Throughput & 6.4 GB/s & 800 MB/s \\
        Frequency & 800MHz & 100MHz \\
        Compute units & $256\times256$ &  $32\times32$  \\
        Input Scratchpad & $6144$ KB & $512$ KB \\
        Weight Scratchpad & $6144$ KB & $512$ KB \\
        Output Scratchpad & $2048$ KB & $192$ KB \\
        \hline
    \end{tabular}
    }
    \caption{\textbf{\sys{} design parameters used in the analytical model and the cycle-accurate simulator.}}
    \vspace{-2em}
    \label{tbl:exp_param}
\end{table}

The ML inference accelerator is a decoupled-access-execute (DAE) unit with a systolic array implementation. 
Such accelerators are widely used in cloud~\cite{tpuv4,AWS-Graviton2} and in edge devices~\cite{VTA,eyeriss}.
These accelerators have a fixed dataflow to simplify the hardware design. 
We choose an output-stationary dataflow, used in Google TPUs in our implementation, but \sys{} can also be used for other dataflows~\cite{eyeriss,maeri}.
The \textit{crypto} unit shown in ~\cref{fig:sec_arch} use parallel AES-GCM implementation from prior work~\cite{aespll}. 
The integrity counters are generated inside the accelerator, as implemented by mgx~\cite{mgx2022isca}.
The secure design also includes a constant traffic \textit{shaper}, whose bandwidth is programmed during tenant bootstrapping as proposed by sesame~\cite{sesame}. 
It also includes a hardware \textit{zeroizer} implementation, taken from sesame.
The input write scratchpad port is multiplexed between the main data path and the \textit{zeroizer} unit.

We evaluate \sys{} on two secure accelerator designs, with configurations listed in ~\cref{tbl:exp_param}.
The first configuration simulates a cloud MLaaS deployment similar to a Cloud TPU-v1~\cite{TPU}, while the second smaller configuration simulates an edge deployment, with configuration similar to Eyeriss-v2 accelerator~\cite{eyeriss}.
We use a pipelined \textit{crypto} unit~\cite{aespipe} having 1 cycle encryption and a 1 cycle integrity latency.
Multiple of these units are instantiated to achieve a crypto throughput listed in ~\cref{tbl:exp_param}.


\subsection{Analytical model implementation}
\label{subsec:impl_anal}
The layer tiling and loop order exploration is performed by the gamma~\cite{gamma} mapping tool and the latency and the energy consumption is estimated with the maestro~\cite{maestro} cost model.
The datasize of individual data is 32 bits.
The dataflow is set to output stationary (os) and the offchip bandwidth is set to the \textit{crypto} throughput mentioned in ~\cref{tbl:exp_param}.
The energy consumed for each decryption and integrity verification is taken from prior work~\cite{aespll}.
These numbers are multiplied with the activity count of the encryption and the integrity units.
The hash granularity is tuned from $64B$ to $4kB$ in powers of 2.
We implement a memory trace generator, that takes mapping loop-order and tiles to generate a demand traffic sequence for each mapping.
The trace generator inserts the redundant transactions and integrity hash memory requests, while exploring the optimal hash granularity.
The final memory trace is used to estimate the energy consumed during data transfers. 
The energy cost of each read and write request is taken from DRAMPower~\cite{drampower}.
The peak occupancy of individual ifmap, weight and ofmap scratchpads for different mappings provide the zeroization latency. 
The unit latency and energy cost for zeroization is the same as scratchpad write operation in maestro.
The analytical model first generates \textit{top-k} mappings for each layer, which is fed to the simulated annealing algorithm to generate 
\textit{top-m} mappings for the entire model. 
The algorithm is run for 1000 iterations with m parallel instances.

\subsection{Simulator implementation}
\label{subsec:impl_sim}
\sys{} uses the cycle-accurate scale-sim~\cite{scalesim} simulator to evaluate the compiler optimizations.
The simulator is augmented to include the secure hardware blocks and also an instruction queue.
The model binary is loaded in the instruction queue. 
The \texttt{load} and \texttt{store} transactions are sent to the demand queue of the traffic shaper. 
There is a separate \textit{shaper} logic for the read and the write bus, with individual demand queues.
These queues are of finite size and stalls the system during times of high demand. 
The \textit{shaper} also sends a fake transaction every time the demand queue is empty. 
The \texttt{compute} instructions are issued to the PE for computation and the \texttt{zeroize} instructions are sent to the \textit{zeroizer}.
The \textit{crypto} block is simulated with a constant latency added to each memory transaction at the appropriate granularity.
The \textit{zeroizer} block shares the scratchpad write port with the main data pipeline. 
It contains a queue for holding zeroize instructions, similar to the sesame implementation and zeroizes 256 bytes in scratchpad every cycle. 
The main data path has a higher priority than the zeroize instructions.
The DRAM transactions are simulated with the ramulator~\cite{ramulator} tool.
The arrival time of each transaction is added to the simulator tick of request dispatch.
The DRAM follows a close-row policy to eradicate timing attacks from row buffer hits. 

The simulator is fed with a model binary, generated by the compiler. The \texttt{load} and \texttt{store} instructions are loaded into the respective traffic shaper demand queue. The \texttt{gemm} instructions are simulated by the compute unit, and the \texttt{zeroize} instruction zeroizes specific regions of a scratchpad. 

\subsection{Benchmarks}
\label{subsec:impl_bench}
Benchmarks are chosen to evaluate different types of ML models.
This includes applications like image classification -- \textbf{Alexnet}~\cite{alexnet} and \textbf{Resnet-50}~\cite{resnetv2};
object detection -- \textbf{FasterRCNN}~\cite{fasterrcnn},
recommendation systems -- \textbf{DLRM}~\cite{DLRM},
a gaming bot -- \textbf{AlphaGoZero}~\cite{AlphaGoZero},
the encoding layer of a LSTM language translation model -- \textbf{Translate}~\cite{translate},
a basic transformer with text embeddings -- \textbf{Transformer}~\cite{Transformer}.
\section{Evaluation}
We present the benefit of the analytical exploration phase in~\cref{subsec:eval_anal}.
Next, \cref{subsec:eval_compiler} shows how the compiler phase reorders load transactions to further improve the runtime of
each mappings.
Finally, \cref{subsec:zero_eval} evaluates the effectiveness of proactive zeroization in a time-shared multi-tenant environment.

\subsection{Analytical exploration phase}
\label{subsec:eval_anal}
We showcase the runtime latency reduction across different mapping phases of \sys{} analytical model.
The results of each analysis is normalized against a baseline mapping
($base_{map}$).
The b $base_{map}$ configuration uses an unoptimized mapping.
We take the top mapping after running the tiling exploration for one generation, with an integrity granularity of $64B$, 
and a constant traffic bandwidth set by the user. 

\textbf{Tiling and loop order exploration:}
This stage provides an initial approximation of the best layer mappings.  
We run the gamma mapper with runtime latency and energy consumption as our fitness function for multiple generations. 
The algorithm stops when the improvement of the best mapping is <$5\%$ compared to the last two generations. 
Convolution layers with a small \textit{ifmap} and \textit{filter} sizes, found in Resnet50, FasterRCNN and transformer converge, after running <$100$ generations. 
Large convolution layers found in alexnet and DLRM takes 100-300 generations. 
The embedding layer of translate and transformer and some layers in resnet50 and FasterRCNN with a large number of channels take >$1000$ generations.
A couple of large convolutional layers of DLRM and AlphaGoZero also fails to converge and we see a large performance variation in the mappings.
For these runs, we terminate the run after 1000 generations. 
\begin{figure}[htbp]
    \centering
    \includegraphics[width=\linewidth]{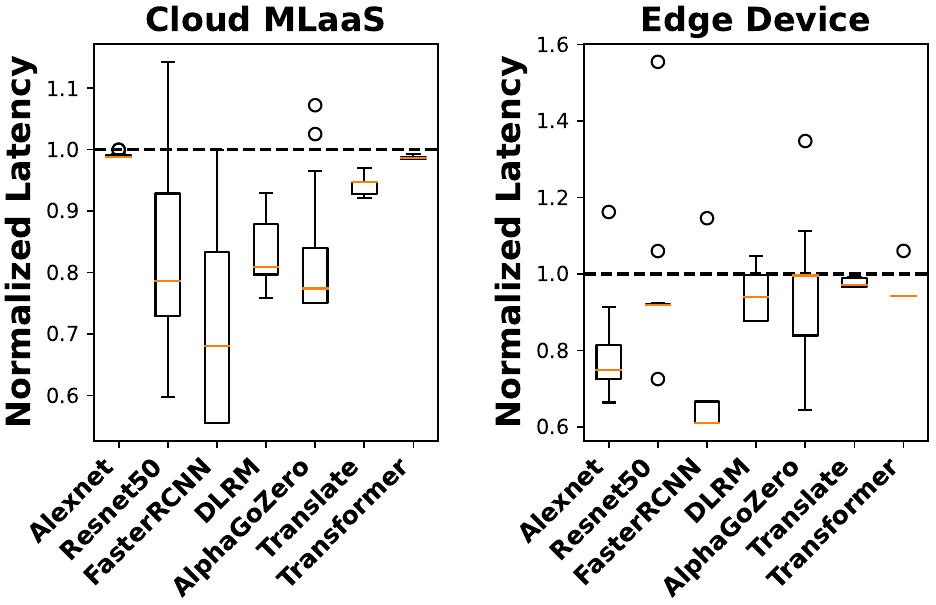}
    \caption{ \textbf{The boxplot shows the performance of 20 model mappings with layer maps chosen from the \\\textit{top-k} output of the tiling and loop order exploration.}}
    \vspace{-1em}
    \label{fig:gamma_dse}
\end{figure}
~\cref{fig:gamma_dse} shows the performance variance of 20 randomly chosen mappings of each model normalized against $base_{map}$.
To show the performance variance, we derive the best mapping by taking the top mapping from each layer.
The worst model mapping is derived by choosing the bottom mapping of each layer. 
We see an average latency improvement of $22.5\%$ in cloud MLaaS and 
$26.2\%$ improvement for the edge deployment for the latency critical layer. 
There is a larger latency variance in the cloud MLaaS deployment than the edge device. 
This is due to more tiling options created by a larger accelerator. 
Some of the configurations are worse than the $base_{map}$. 
This is due to the fact that we sort the genetic population after the completed generation to extract the \textit{top-k} mappings.
The bottom mappings of some of the layers perform worse than the top configuration after the first generation (used in $base_{map}$). 
These configurations will easily be rejected by the simulated annealing algorithm.

\begin{figure}[t]
    \centering
    \includegraphics[width=\linewidth]{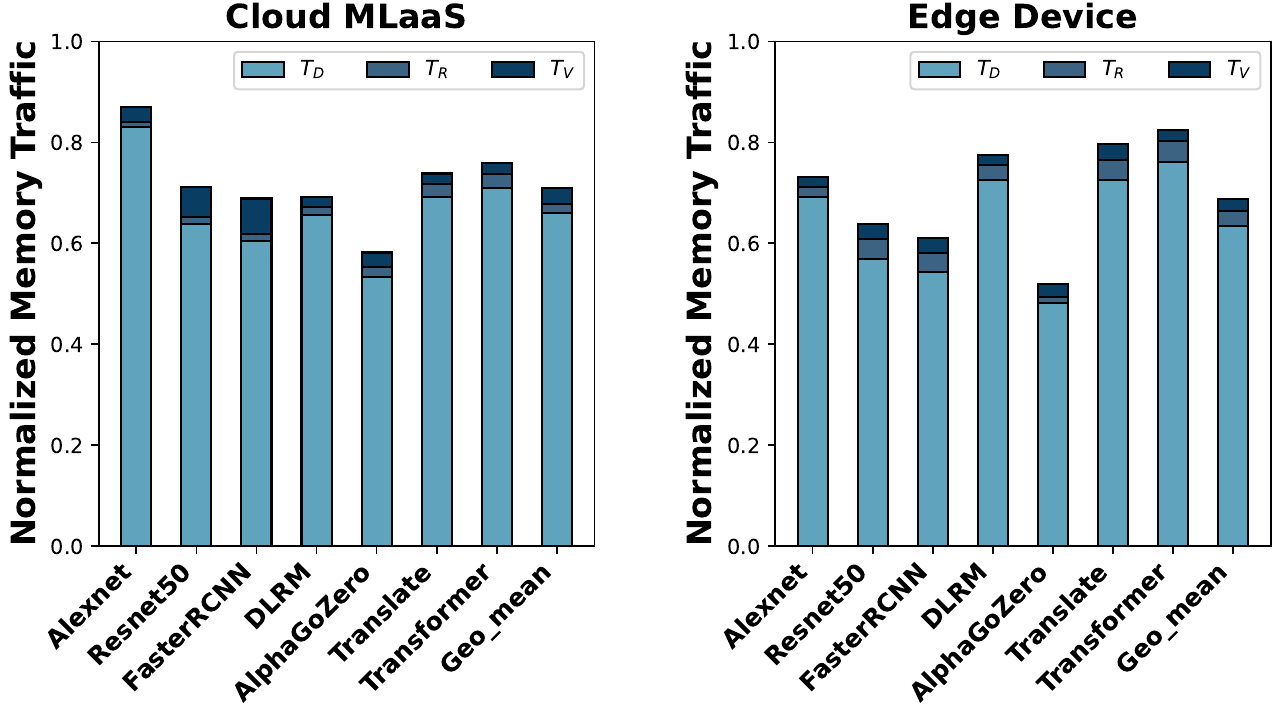}
    \vspace{-2em}
    \caption{ \textbf{\textbf{Memory traffic reduction after hash granularity optimization. $\{T_D,T_R,T_V\}$ 
    are demand, redundant and hash memory requests.
    }}}
    \vspace{-1em}
    \label{fig:gamma_int_opt}
\end{figure}

\textbf{Exploring the optimal integrity granularity:}
Once we generate the top-k mapping for each layer, we explore the optimal integrity granularity.
The data sizes considered for hash computation is varied from $64B$ to $4kB$ stepping by a power of 2. 
The optimization target is to find the setting with the minimum number of memory transactions. 
~\cref{fig:gamma_int_opt} compares the memory traffic of the \textit{top-1} configuration normalized with the $base_{map}$ traffic. 
The memory traffic is computed by adding the optimal hash mapping for each layer, having the minimum number of memory requests.
There is an average memory traffic reduction of $29.1\%$ for cloud MLaaS and $31.2\%$ for edge deployment. 
Large layers in AlphGoZero benefit from a larger hash granularity. 
Same is true for layers with large number of channels found in Resnet50 and FasterRCNN. 
This traffic reduction saves system energy and memory congestion, brought by an efficient data reuse for each layer,
along with less memory integrity traffic.

\begin{figure}[htbp]
    \centering
    \includegraphics[width=\linewidth]{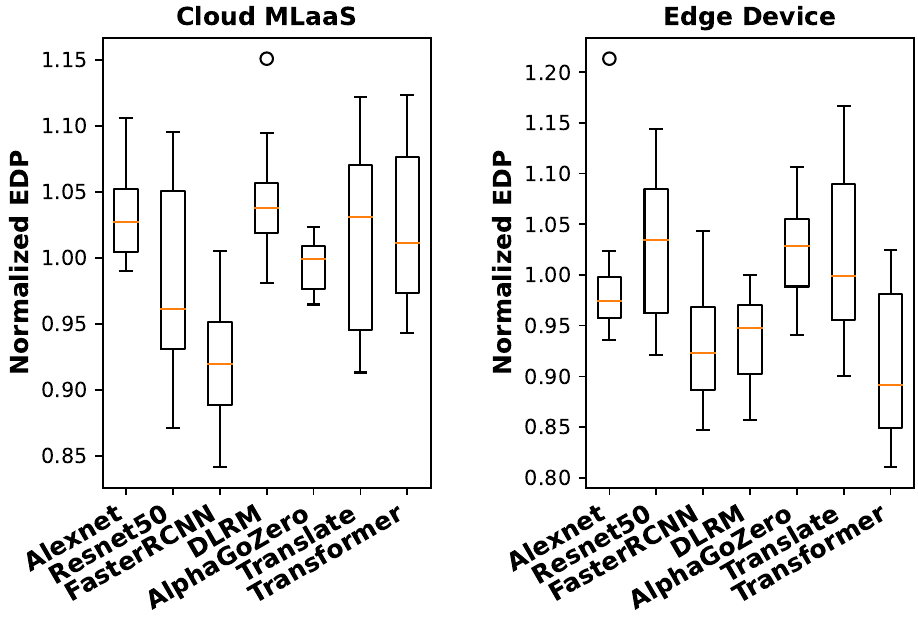}
    \vspace{-2em}
    \caption{ \textbf{\textbf{EDP of \textit{top-m} model mappings generated by simulated annealing.
    The result is normalized against the model map with all the top layer mappings.
    }}}
    \vspace{-1.5em}
    \label{fig:se_topm}
\end{figure}

\textbf{Finding the model mappings:}
We find the top-m model mappings by the simulated annealing (SA) algorithm. 
The top-m mappings uses energy-delay product as the optimization target. 
\cref{fig:se_topm} compares the EDP of the \textit{top-m} model mappings normalized against the \textit{top-1} configuration generated by the tiling phase.
SA is able to find better mappings from the search space in all cases. 
The reason is the fact that SA chooses an optimal \textit{shaper} bandwidth with the EDP objective. 
This leads to an EDP improvement of upto $15\%$ in cloud and $20\%$ in edge compared to the layer mapping outputs.

\subsection{Compiler optimizations}
\label{subsec:eval_compiler}
\subsubsection{Performance and energy improvements}
\label{subsec:perf_compiler}
\begin{figure}[t]
    \centering
    \vspace{-0.5em}
    \includegraphics[width=\linewidth]{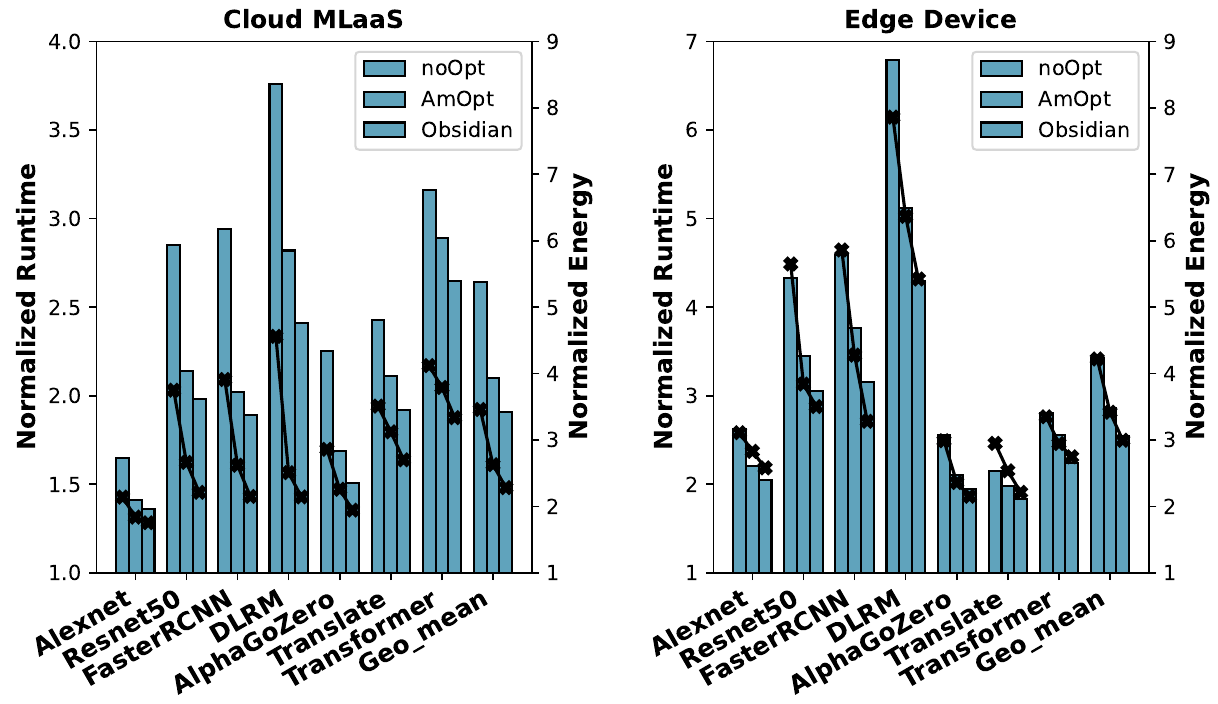}
    \vspace{-2em}
    \caption{ \textbf{\textbf{Runtime and energy consumption comparison among three configurations.
    The best bar is a non-optimized mapping, the second one is the top-1 mapping from analytical model, 
    while the last one additionally includes the compiler optimizations.
    }}}
    \vspace{-1.5em}
    \label{fig:perf_edp}
\end{figure}

\begin{figure}[b]
    \centering
    \includegraphics[width=0.95\linewidth]{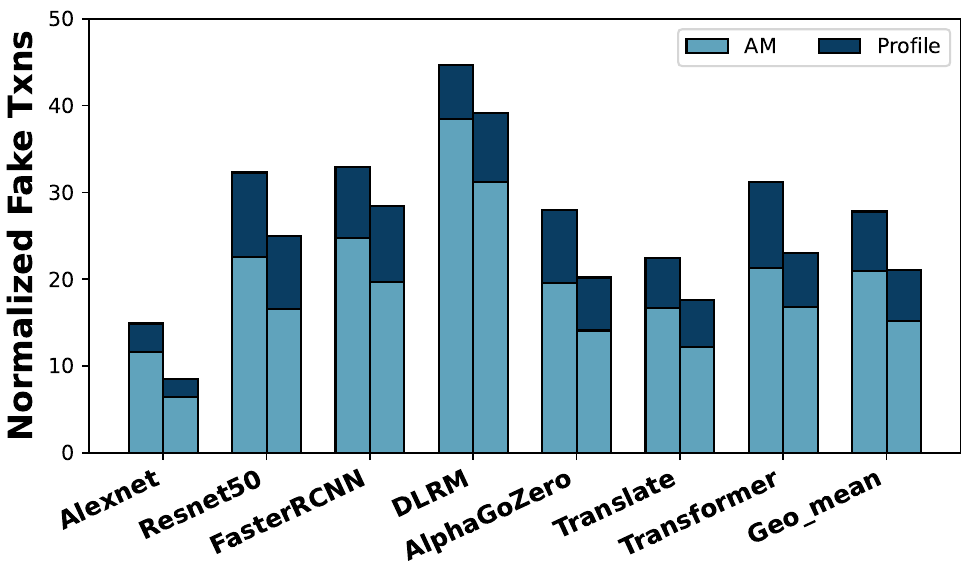}
    \vspace{-1em}
    \caption{ \textbf{\textbf{Fake transaction reciprocated by the compiler by promoting data loads from subsequent layers. }}}
    \vspace{-1em}
    \label{fig:fake_txn}
\end{figure}
\vspace{-0.5em}
We further tune the top-m mappings in a cycle-accurate simulator described in ~\cref{subsec:impl_sim} to find the optimal mapping. 
We profile each configurations to promote more load transactions
to reciprocate fake transactions into successful prefetches. 
~\cref{fig:perf_edp} compares the performance and the energy consumption of secure ML executions normalized to a non-secure baseline. 
The mapping for each layer of the non-secure baseline is generated by the tiling exploration stage with memory bandwidth set in ~\cref{tbl:exp_param}. 
The bars show the normalized runtime, while the line plot shows the normalized energy consumption. 
The first bar shows the runtime of the $base_{target}$. For brevity, we call this configuration as \textbf{noOpt}.
The second bar is the top configuration generated by the analytical model (\textbf{AmOpt}).
The third bar represents the \sys{} configuration with both analytical and compiler optimizations (\textbf{Obsidian}).
The \textbf{AmOpt} configuration brings down both the latency by $20.5\%$ for the cloud and $8.4\%$ for the edge deployment. 
The cloud configuration presents a larger state space and more optimization opportunities for the analytical model compared to a constrained edge deployment. 
There is also an energy improvement of $24\%$ for cloud and $8.4\%$ for the edge. 
This is due to the tile mapping creating a greater synergy between the memory and the PE throughput,
an optimal hash granularity and traffic shaper bandwidth selection. 

The compiler phase, further improves the cloud latency and energy by $9.1\%$ and $13.8\%$,
and the edge latency and energy by $12.2\%$ and $13.1\%$. 
The edge deployment, with its more constrained resources, benefits more from compiler optimizations.
The edge execution has multiple bottlenecks, which are relaxed by store to load forwarding as well as 
from load promotion. 

\begin{figure}[t]
    \centering
    \includegraphics[width=\linewidth]{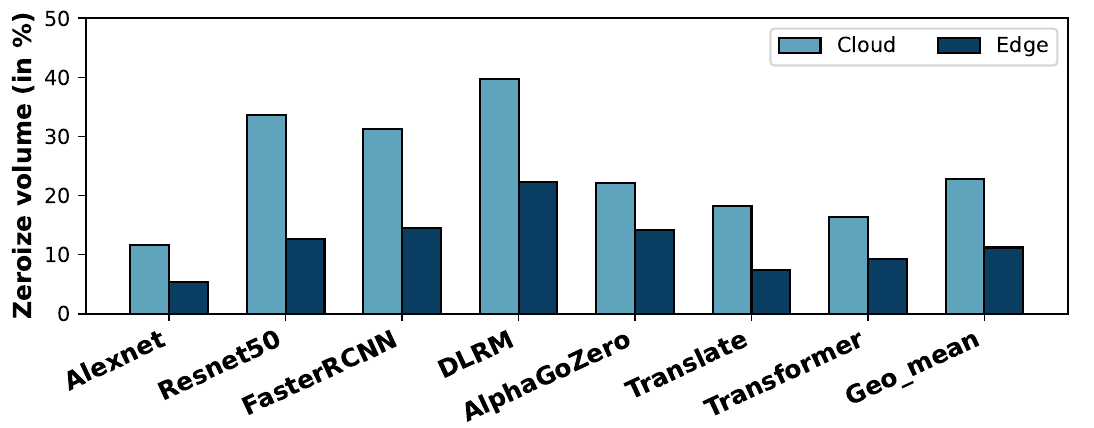}
    \vspace{-2em}
    \caption{ \textbf{\textbf{Data volume zeroized by proactive zeroization with layer computation for cloud and edge devices.}}}
    \vspace{-1em}
    \label{fig:zeroize}
\end{figure}
Overall, \sys{} provides an average speedup of $29.6\%$ in cloud and $20.6\%$ in edge deployments over the $base_{map}$ configuration.
It also provides an energy improvement of $37.8\%$ for cloud MLaaS and $21.5\%$ for edge devices. 
\sys{} provides a faster secure ML inference, with energy efficiency in both datacenters and 
battery-operated edge devices.


\subsubsection{Reciprocation of fake transactions}
We deep dive into the real cause of the performance and energy improvement from compiler profiling described in ~\cref{subsec:perf_compiler}.
One of the main reason is replacement of fake transactions generated by the \textit{shaper} with useful requests. 
~\cref{fig:fake_txn} shows the reduction in the number of fake transaction in the overall execution compared to $base_{map}$. 
The two plots show the cloud and edge results respectively.
The \textbf{AM} region of each bar shows the fake traffic reduction by the analytical model by efficient state space exploration. 
The \textbf{Profile} region shows how the compiler passes bring down the fake transaction, reciprocating it with useful memory traffic. 
The low fake transaction reciprocation for alexnet, as large layers will up the scratchpad reducing load promotion opportunities.
The high reciprocation for DLRM, due to its small filter size and compute intensive layers. It creates greater chance to load subsequent model weights into the scratchpad. 
The smaller layers found in Resnet50 and Fasterrcnn lead to more than $20\%$ fake transaction reciprocation.

\vspace{-0.5em}
\subsection{Proactive zeroization}
\label{subsec:zero_eval}

We consider a time-shared multi-tenant accelerator, where models are context switched at each layer boundary.
The compiler liveness analysis zeroizes stale data in each scratchpad, in parallel to the computation. 
In this case, cross layer load promotion is disabled as each consecutive model layer is interleaved with other tenants.
The zeroize volume in~\cref{fig:zeroize} is defined by the percentage of secret data zeroized proactively during layer execution. 
The cloud execution involves larger tiles with higher compute duration, presenting a higher opportunity for proactive zeroization than edge devices.
The liveness analysis zeroizes, $22.8\%$ data volume for cloud and $11.2\%$ for the edge deployment.
DLRM layers are compute bound with larger tiles, which provide more opportunity to interleave zeroization with computation. 
Same is observed for certain Resnet50 and FasterRCNN layers with a large channel size, which led to larger tile mappings.

\vspace{-.5em}
\section{Related works}

\noindent {\bf State space exploration of ML accelerators.} 
Timeloop~\cite{timeloop} and maestro~\cite{maestro} build analytical models to explore the large design space and show the benefits of loop ordering and tensor tiling operations. 
However, these models do not model the security primitives.
Secureloop~\cite{secureloop} builds on timeloop to find the optimal authentication block. 
While they perform a more exhaustive search to find the optimal hash block,
it is not clear how their approach extend to other 
security primitives (traffic shaper and zeroization logic).
Another line of work~\cite{AutoTVM,flextensor} uses compiler profiling for state space exploration. 
However, these works are also limited to non-secure accelerators. 

\noindent {\bf Secure ML accelerator architectures.}
The growing importance of privacy preserving ML inference in cloud and edge has led to several 
secure accelerator trusted execution environment designs~\cite{GuardNN2022DAC,sesame,securator,keystone,triton}. 
Most of these designs focus on optimizing the \textit{crypto} block. 
For instance, TNPU, GuardNN, and MGX~\cite{tnpu2022HPCA,GuardNN2022DAC,mgx2022isca} proposed tree-free integrity verification
exploiting DNN-specific data access patterns. 
We leverage this optimization in our design.
Sesame additionally includes a constant traffic shaper and zeroizer to protect model hyperparameters and the scratchpad memory safety. 
\sys{} focuses on improving the performance and energy consumption of secure accelerators through state-space exploration.
A group of work~\cite{Secureml,Cryptonets} use homomorphic encryption and multi-party computation to secure accelerators, but these methods are orthogonal to the TEE based design used in this work. 

\noindent {\bf Multi-tenancy in ML accelerators.}
Prema and AI-MT~\cite{prema,AIMT2020ISCA} explored preemptive scheduling algorithms for temporal
multi-tasking on accelerators.
Both explored interleaving memory and compute operations to improve efficiency.
MAGMA~\cite{magma2022HPCA} explored optimization algorithms for mapping
multi-tenant workloads onto accelerators.
While, tenant scheduling is orthogonal to this work,
zeroization is necessary to protect memory safety during context switch.
\sys{} optimizes the \textit{zeroizer} to reduce context switch latency.

\section{Conclusion}
\sys{} explores the workload state-space exploration in a secure accelerator. 
The analytical and the compiler profiling approach has its own set of problems. 
While the analytical model explores a larger state space, the compiler profiling takes runtime execution
bottlenecks into account. 
\sys{} uses a combination of these approaches to find the optimal mapping. 
By leveraging the large state space exploration capability of an analytical model, it reduces the space to 
a smaller number of mappings to be profiled in a compiler. 
This approach finds workload mappings that greatly reduce the runtime latency and the energy consumption of a secure ML accelerator inference.

\bibliographystyle{plain}
\bibliography{obsidian_hw_main.bbl}
\end{document}